\documentstyle[floats,epsf,aps,prd]{revtex}

\def\Journal#1#2#3#4{{#1} {\bf #2}, #3 (#4)}

\def\NP{{Nucl.~Phys.}}
\def\PL{{Phys.~Lett.}}
\def\PR{{Phys.~Rev.}}

\def\PRD{{Phys.~Rev.~D}}
\def\PRL{Phys.~Rev.~Lett.}
\def\ZP{{Z.~Phys.}}
\def\PTP{{Prog.~Theor.~Phys.}}

\newcommand{\bra}[1]{\langle #1 |}
\newcommand{\ket}[1]{|#1\rangle}

\newcommand{\mn}{{\mu\nu}}
\newcommand{\ab}{{\alpha\beta}}
\newcommand{\tr}{{\rm tr}}

\newcommand{\T} {\mbox{T}}

\def\J{$J/\psi$}

\title{%
{\normalsize RBRC-56 \hfill BNL-66659}\\
Long--Range Forces of QCD}

\author{%
   H.~Fujii\thanks{Present address: Institute of Physics, University
   of Tokyo, Komaba, Meguro, Tokyo 153-8902.} and D.~Kharzeev}

\address{
   RIKEN-BNL Research Center,\\
   Brookhaven National Laboratory,\\ 
Upton, NY 11973, USA }
\date{\today}
 
\begin{document} 
\draft
\maketitle

\begin{abstract}
We consider the scattering of two color dipoles (e.g., heavy
quarkonium states) at low energy -- a QCD analog of Van der Waals
interaction. Even though the couplings of the dipoles to the gluon
field can be described in perturbation theory, which leads to the
potential proportional to $(N_c^2-1)/R^{7}$, at large distances $R$
the interaction becomes totally non-perturbative. Low--energy QCD
theorems are used to evaluate the leading long--distance contribution 
$\sim (N_f^2-1)/(11N_c - 2N_f)^2\ R^{-5/2}\ exp(-2 \mu R)$ 
($\mu$ is the Goldstone boson mass), which is shown to arise from the
correlated two--boson exchange. The sum rule which relates the overall
strength of the interaction to the energy density of QCD vacuum is
derived. Surprisingly, we find that when the size of the dipoles
shrinks to zero (the heavy quark limit in the case of quarkonia), the
non-perturbative part of the interaction vanishes more slowly than the
perturbative part as a consequence of scale anomaly. As an
application, we evaluate elastic $\pi J/\psi$ and $\pi J/\psi \to \pi
\psi'$ cross sections.      
\end{abstract}

\section{Introduction}

The interaction between small color dipoles\footnote{Small color
dipoles can be realized in the real word as heavy quarkonium states or
as virtual quark--antiquark pairs in deep inelastic scattering.}
provides an interesting theoretical laboratory for the studies of QCD
and its applications in nuclear physics. Indeed, the asymptotic
freedom dictates that the coupling of strong interactions becomes weak
at short distances, and since the small size of dipoles introduces a
natural infrared cut-off, one can hope that their interactions can be
systematically treated in perturbation theory
\cite{AF,A,Gottfried,Voloshin,Peskin,Leut}.

One could therefore expect that at low energy the interaction
between the dipoles in $SU(N)$ gauge theory would be of Van der Waals
type:  
\begin{equation}
V_{\rm pert}(R) \sim - g^4 (N^2-1)\ {1 \over R^n}, \label{vdw}
\end{equation}
where $n=6$ in the original Van der Waals potential, and $g$ is the
gluon coupling evaluated at the scale of quarkonium size. Indeed, this 
behavior was established by Appelquist and Fischler 
\cite{AF}, who studied the interactions of static color dipoles 
described by Wilson loops. These authors also explored the 
breakdown of the perturbative expansion in the static potential \cite{A}, 
and pointed to the possibility that retardation effects can modify 
the $1/R^6$ dependence once the spatial motion of the quarks is considered.
In this paper, we take this effect into account and argue that, in the
limit of the small size of the dipoles, the potential (\ref{vdw}) is
actually of Casimir-Polder \cite{cas} type, with $n=7$.
On the other hand, gluons cannot propagate at large distances, where
the dominant degrees of freedom are the lightest hadronic states. In
the chiral limit, the theory with spontaneously broken 
$SU_L(N_f) \times SU_R(N_f)$ symmetry contains $(N_f^2-1)$ Goldstone
bosons, and the number of flavors $N_f$ should effectively replace the
number of colors in the coefficient of Eq.~(\ref{vdw}) at large
distances:
\begin{equation}
V_{\rm chiral}(R) \sim - (N_f^2-1)\ {1 \over R^n}. \label{vdwc}
\end{equation}
In the real world where the masses $\mu$ of Goldstone bosons are not equal to 
zero, instead of Eq.~(\ref{vdwc}) at large distances one expects to find 
the potential of Yukawa form
\begin{equation}
V(R) \sim - (N_f^2-1)\ {e^{- 2 \mu R} \over R}. 
\label{vdwy}
\end{equation}
(We will show that the actual form of the long--distance potential is
different from (\ref{vdwy}) -- see Eq. (\ref{uncor}).) How does the
transition between the behavior at short and long distances occur? Can
one explicitly, from the first principles, evaluate the long--distance
potential?

In this paper we address these questions, and argue that the
interaction between small color dipoles (heavy quarkonium states in
our example) at large distances can be reliably evaluated. Our
analysis is based on the following two properties of QCD:  
1) the scale invariance which is present at the tree level in QCD 
with massless quarks is broken by interactions; this is reflected in
the non-zero divergence of scale current, and hence non-vanishing
trace of the energy-momentum tensor \cite{scale,collins}; 
2) the chiral symmetry is broken spontaneously, which implies the
existence of Goldstone bosons; being the lightest of all hadrons, they
are the relevant degrees of freedom at large distances.

These two properties of QCD are beautifully linked by the low--energy
theorem derived by Voloshin and Zakharov \cite{VZ}, which we discuss
below. The first of these properties was previously exploited to
derive the low--energy amplitude of quarkonium--nucleon scattering
\cite{AK,LMS} (for recent work, see \cite{Hay,Hood}). Van der Waals
interactions of quarkonium with nucleons and nuclei were discussed in
Refs.~\cite{Bro,DW,C,BM}.
For applications to the low--energy quarkonium--pion scattering and
the structure of quarkonium, see \cite{DK,SSZ,CS,GR}. Quarkonium
dissociation cross sections in interactions with light hadrons were
evaluated in Refs.~\cite{Peskin,Kai,KS}. Some of the results of this
study were previously reported in Ref.\cite{FK}.

The picture which emerges from our approach is the following. The
heavy ``onia'' couple perturbatively to the gluon field; at small
distances, the entire interaction can be evaluated perturbatively. At
larger distances, however, the interaction becomes grossly modified by
the coupling to pion fields, which is fixed by low-energy
theorems. The dominance of the non-perturbative interaction at large
distances in this case will be shown to be a consequence of the finite
energy density of QCD vacuum \footnote{This picture was foreseen by
Bjorken \cite{Bjorken}.}. 
We also find that when the size of the dipoles shrinks to zero (which
is what happens in the heavy quark limit with quarkonia), the
non-perturbative part of the interaction vanishes more slowly than the
perturbative part -- in other words, the interaction between very
small dipoles becomes totally non-perturbative! This surprising result
will be shown to be a natural consequence of scale anomaly in QCD.

In this paper, we limit ourselves to the interaction at small
energies; however we hope that some of our results may be extended to
the case of dipole scattering at high energies \cite{Mueller}, where
the broken chiral symmetry can also play a substantial role, as
discussed by Anselm and Gribov \cite{AG}.

\vskip0.3cm
The paper is organized as follows. In Section \ref{sec:pert}, we give
the general expression for the scattering amplitude of two color
dipoles in the framework of the Operator Product Expansion (OPE),
introduce the spectral representation method for the evaluation of
this amplitude, and use this method to re-derive the perturbative
expression \cite{Peskin} for the low--energy scattering amplitude (or
potential). In Section \ref{sec:beyond}, we discuss the scattering
amplitude of color dipoles beyond the perturbation theory, derive the
leading long--distance behavior of the potential, and discuss the
relative strength of perturbative and non-perturbative
contributions. In Section \ref{jpsipot}, we evaluate the potential
acting between two $J/\psi$'s. In Section \ref{sec:sumrule} we use the
low energy theorems \cite{VZ,NS,MS,nsvz} to derive the sum rule
relating the strength of the potential to the energy density of QCD
vacuum. In Section \ref{sec:pion}, we evaluate the cross sections of
\J\ interactions with pions, relevant for the problem of \J\
suppression in heavy ion collisions \cite{MS1,QM}. The final Section
\ref{sec:sum} is devoted to summary and discussion.

\section{Interaction of Color Dipoles in Perturbation Theory}
\label{sec:pert}

The small size of the heavy quarkonium $\Phi$ allows us to expand the
amplitude of its interaction with hadrons ($h, h'$) at low energy in
the form of multipole, or operator product,
expansion\cite{Voloshin,Peskin}: 
\begin{eqnarray}
{\cal M} = \sum_i c_i \bra{h'}O_i(0)\ket{h} ,
\label{ope0}
\end{eqnarray}
where $O_i(x)$ are the gauge-invariant local operators and   
$c_i$ are the Wilson coefficients (polarizabilities) which reflect the
structure of the quarkonium; the energy of the hadrons is assumed to
be small compared to the binding energy of the quarkonium,
$\epsilon_0$. The factorization scale in this formula can be chosen at
$\epsilon_0$. At small energies, the leading operator in
Eq.~(\ref{ope0}) is the square of the chromo-electric  
field $(1/2) g^2 {\bf E}^{a2}(0)$ \cite{Voloshin,Peskin} -- this is the
leading twist--two operator expressible in terms of gluon fields.
Other twist-two operators contain covariant derivatives leading to the
powers of the ratio of the energy transfer to the binding energy and
are therefore suppressed at small energy;  
the series of these local operators can be summed up into a double
dipole form\cite{Peskin}: 
\begin{eqnarray}  
\frac{g^2}{N} \sum_{n=2, {\rm even}}^\infty
\bra{\phi}r^i\frac{1}{(H_a+\epsilon)^{n-1}}r^j\ket{\phi} \ 
\tr[E_i(0) (-iD^0)^{n-2} E_j(0)] 
&=&
\frac{g^2}{N} \langle 
\tr \Big [ {\bf r\cdot E}(0) \frac{1}{H_a+\epsilon +iD^0}{\bf r\cdot E}(0)
\Big ] \rangle \ ,
\label{eq5}
\end{eqnarray}
where the Wilson coefficients are explicitly given by the expectation 
values over the singlet state $\phi(r)$ with the binding energy $\epsilon$;
$H_a(r)$ is the effective Hamiltonian describing the intermediate,
$SU(N)$ color-adjoint quark-antiquark state;  
$D^0$ is the covariant derivative acting on {\bf E}
and the trace over the color indices of gluon operators
ensures that Eq.~(\ref{eq5}) is gauge-invariant. 
In the heavy quark limit, $\phi(r)$ can 
be approximated by the Coulomb wave function.

Using this multipole representation, one can write
down the amplitude of the scattering of two small
color dipoles at low 
energies (in the Born approximation) in the following form\cite{Peskin}:
\begin{eqnarray}
V(R) &=& -i \int_{-\infty}^\infty dt\   
\bra{0} \T \big( \sum_i c_i O_i (x)\big) 
           \big( \sum_j c_j O_j (0)\big) \ket{0} . 
\label{ope1}
\end{eqnarray}
Keeping only the leading operators, we can rewrite Eq.\ (\ref{ope1})
in a simple form,
\begin{eqnarray}
V(R) &=& 
- i\Big ( \bar d_2 \frac{a_0^2}{\epsilon_0} \Big)^2 
\int_{-\infty}^\infty dt\   
\bra{0} \T \ \frac{1}{2}g^2 {\bf E}^{a}\cdot{\bf E}^{a} (t,R)\ 
             \frac{1}{2}g^2 {\bf E}^{b}\cdot{\bf E}^{b}(0) \ket{0}, 
\label{pot1}
\end{eqnarray}
where $\bar d_2$ is the corresponding Wilson coefficient defined by
\begin{equation}
\bar d_2 \frac{a_0^2}{\epsilon_0}
=\frac{1}{3N}\bra{\phi} r^i \frac{1}{H_a + \epsilon} r^i \ket{\phi}, 
\label{wilson}
\end{equation}
from which we 
have explicitly factored out the dependence on the quarkonium Bohr 
radius $a_0$ and the Rydberg energy $\epsilon_0$.  The factors 
$a_0$ and $1/\epsilon_0$ represent the characteristic dimension
and fluctuation time of the color dipole, respectively. 
The approximation used in deriving Eq.~(\ref{pot1}) is justified 
when the gluon fields
change slowly compared to $1/\epsilon_0$. The Wilson coefficients 
(\ref{wilson}) were computed for S \cite{Peskin} and P \cite{DK} 
states of quarkonium in the large $N$ limit.

In physical terms, the structure of Eq.\ (\ref{pot1}) is transparent: 
it describes elastic scattering of two dipoles which act on 
each other by chromo-electric dipole fields; color neutrality permits 
only the square of dipole interaction.

The amplitude (\ref{pot1}) was evaluated before \cite{Peskin} 
in perturbative QCD using functional methods. 
For our purposes, however, it is convenient to use a spectral
representation approach\footnote{The use of dispersion theory in
electrodynamics for the interaction between neutral atoms was
pioneered by Feinberg and Sucher \cite{FS}.}. 
As a first application of this approach, we will reproduce the result
of \cite{Peskin} by a different, and perhaps more simple, method.

First, it is convenient to express $g^2 {\bf E}^{a2}$ in terms of the 
gluon field strength tensor \cite{NS}:
\begin{eqnarray}
g^2 {\bf E}^{a2} &=&
\frac{g^2}{2}({\bf E}^{a2}-{\bf B}^{a2}) 
+ \frac{g^2}{2}({\bf E}^{a2}+{\bf B}^{a2})
\nonumber \\
&=&
- \frac{1}{4}g^2    G^a_{\alpha\beta}G^{a\alpha\beta}
+g^2(- G^a_{0\alpha} G_0^{a\alpha}
+\frac{1}{4} g_{00} G^a_{\alpha\beta}G^{a\alpha\beta})
=
 \frac{8 \pi^2}{b} \theta_\mu^\mu + g^2 \theta_{00}^{(G)}  \ ,
 \label{e2}
\end{eqnarray}
where
\begin{eqnarray}
\theta_\mu^\mu \equiv
\frac{\beta(g)}{2g} G^{a\alpha\beta} G_{\alpha\beta}^{a} =
- \frac{b g^2}{32 \pi^2} G^{a\alpha\beta} G_{\alpha\beta}^{a} 
 \ ,
\quad
\theta_\mn^{(G)} \equiv - G^a_{\mu\alpha} G_\nu^{a\alpha}
+\frac{1}{4} g_\mn G^a_{\alpha\beta}G^{a\alpha\beta}
\ . \label{trace}
\end{eqnarray}
Note that $\theta_\mu^\mu$ is the trace of the energy-momentum tensor 
of QCD in the chiral limit, and as a consequence of decoupling 
theorem \cite{svz2} the $\beta$ function in Eq.~(\ref{trace}) 
does not contain the contribution of heavy quarks ({\it i.e.} 
$b=\frac{1}{3}(11N-2N_f)=9$).

Let us now write down the spectral representation for the correlator 
of the trace of the energy-momentum tensor:
\begin{eqnarray}
\bra{0} \T \theta_\mu^\mu(x) \theta_\nu^\nu(0) \ket{0}
&=&
\int \frac{d^4k}{(2\pi)^3} \rho_\theta (k^2) \theta(k_0)
( e^{-ikx} \theta(x^0) + e^{ikx} \theta(-x^0)) 
\nonumber \\
&=&
\int d \sigma^2 \rho_\theta (\sigma^2) \Delta_F(x;\sigma^2),
\label{spectral}
\end{eqnarray}
where the spectral density is defined by
\begin{equation}
\rho_\theta (k^2) \theta(k_0) =
\sum_n (2\pi)^3 \delta^4(p_n-k)|\bra{n}\theta_\mu^\mu\ket{0} |^2 ,
\label{def_spect}
\end{equation}
the phase-space integral should be understood in
Eq.~(\ref{def_spect}), and
\begin{eqnarray}
i\Delta_F(x;\sigma^2) 
&=&
i \int \frac{d^4k}{(2\pi)^3}\delta(k^2-\sigma^2)\theta(k_0)
(e^{-ikx} \theta(x^0)+e^{ikx} \theta(-x^0))
\end{eqnarray}
is the Feynman propagator of a scalar field in the coordinate space.
Substituting the representation (\ref{spectral}) in Eq.\ (\ref{pot1}), 
the potential can be expressed as a 
superposition of Yukawa potentials
corresponding to the exchange of scalar quanta of mass $\sigma$:
\begin{eqnarray}
V_\theta(R) &=& 
-i \Big ( \bar d_2 \frac {a_0^2}{\epsilon_0} \Big )^2 
\Big ( \frac{4 \pi^2 }{b}\Big ) ^2
\int_{-\infty}^\infty dt 
\int d \sigma^2  \rho_\theta(\sigma^2) \Delta_F(x;\sigma^2)
\nonumber \\
&=& -
\Big ( \bar d_2 \frac {a_0^2}{\epsilon_0} \Big )^2 
\Big ( \frac{4 \pi^2 }{b}\Big ) ^2
\int d \sigma^2  \rho_\theta (\sigma^2)
\frac{1}{4\pi R}e^{-\sigma R} .
\label{yukawa}
\end{eqnarray}

Our analysis so far has been completely general; the dynamics 
enters through the spectral density (\ref{def_spect}).
Let us first evaluate this quantity in perturbation theory, where 
it is given by the contributions of two-gluon states
(see Fig.~1(a)) defined by
\begin{eqnarray}
\rho_\theta^{\rm pt}(q^2) & \equiv  &
\sum (2\pi)^3 \delta^4 (p_1+p_2-q) 
| \bra{p_1 \varepsilon_1 a, p_2\varepsilon_2 b} \theta^\mu_\mu 
\ket{0}|^2 ,
\label{pdens1}
\end{eqnarray}
where the phase-space integral is understood,
as well as the summations over the polarization ($\varepsilon_{1,2}$) 
and color indices ($a, b$) of the two gluons.
The calculation for $SU(N)$ color (see Appendix A) gives
\begin{eqnarray}
\rho^{\rm pt}_{\theta}(q^2) &=& 
\left ( \frac{bg^2}{32\pi^2} \right )^2\frac{N^2-1}{4\pi^2}\ q^4;
\label{pdens2}
\end{eqnarray}
the appearance of $q^4$ dependence in Eq.~(\ref{pdens2}) is of course 
natural from dimensional arguments.
Performing the integration in Eq.\ (\ref{yukawa}) over the invariant mass
$\sigma^2$ from zero to infinity, we get the following result 
($N=3$):
\begin{eqnarray}
V^{\rm pt}_\theta(R) &=&
-  g^4 \Big ( \bar d_2 \frac{a_0^2}{\epsilon_0}\Big)^2
\frac{15}{8\pi^3}\frac{1}{R^7}.
\label{casimir}
\end{eqnarray} 
This result can also be derived by the functional method of Bhanot and
Peskin \cite{Peskin} (see Appendix B). 

Several remarks are in order here:
The $\propto R^{-7}$ dependence of the potential (\ref{casimir}) is a 
classical result known from atomic physics \cite{cas}; as is apparent
in our derivation, the extra $R^{-1}$ as compared to the Van der Waals
potential $\propto R^{-6}$ is the consequence of the fact that the
dipoles fluctuate in time, and the characteristic time of fluctuation
$t \sim \epsilon_0^{-1}$ ($\epsilon_0$ is quarkonium binding energy)  
is small compared to the spatial separation of the ``onia'': $t \ll R$
-- note an explicit integration over time in Eq.~(\ref{yukawa}). 
This illustrates, in a somewhat different way, the original argument of
Voloshin \cite{Voloshin} that the physical picture behind the OPE is
orthogonal to the potential model -- the latter is
based on the assumption of instantaneous interaction, whereas the
former is based on the assumption that the internal frequency of heavy
quarkonium $1/\epsilon_0$ is much higher than the frequency of
external soft fields.    
Retardation effects make questionable the possibility to 
describe the interactions of quarks inside a heavy quarkonium 
by a local potential.
In our case, applying the OPE method, we first average the
interactions with soft gluons over the
quarkonium internal state, which corresponds to the infinite
retardation. 
With the resulting coupling between the quarkonium and the gluons,
the potential description of onium-onium scattering is adequate since
at low energies the {\it relative} motion of heavy quarkonia is slow.
The retardation effects manifest themselves in the modification of the
shape of the potential.


We note that although the matrix element of the operator
$\theta_\mu^\mu$ can in general be non-perturbative, in perturbation
theory $\theta_\mu^\mu$ is of order $g^2$, and accordingly the
potential (\ref{casimir}) has the prefactor $g^4$. Then the second
term $g^2 \theta_{00}^{(G)}$ in Eq.\ (\ref{e2}), which describes the
tensor $2^{++}$ state of two gluons, gives the contribution in the
same order in $g$. Adding this contribution to $V_\theta$ in Eq.\
(\ref{casimir}), we recover the complete result of Ref.~\cite{Peskin} 
\begin{eqnarray}
V^{\rm pt}(R) &=&
- g^4 \left (\bar d_2 \frac{a_0^2}{\epsilon_0} \right )^2
\frac{23}{8\pi^3}\frac{1}{R^7};
\end{eqnarray} 
Note that our $\bar d_2$ is related to the $d_2$ in Ref.\
\cite{Peskin} by $ d_2 a_0 \epsilon_0 = \bar d_2 g^2 $.
This perturbative expression is valid when 
$a_0, 1/\epsilon_0 \ll R \ll \Lambda_{\rm QCD}^{-1}$.

\begin{figure}[tb]
\epsfxsize=0.5\textwidth
\centerline{\epsffile{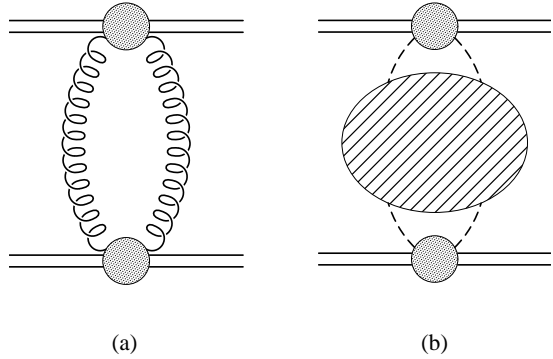}}
\caption{Contributions to the potential between quarkonia from (a) 
two--gluon exchange and (b) correlated two--pion exchange.}     
\end{figure}

\section{Beyond the Perturbation Theory: The Role of Goldstone Bosons}
\label{sec:beyond}

At large distances, the perturbative description breaks down, because 
the potential becomes determined by the spectral density at small $q^2$, 
where the transverse momenta of the gluons become small.

\subsection{Broken scale invariance}

To see the importance of non-perturbative effects explicitly, let
us consider the correlator of $\theta_\mu^\mu$,
\begin{equation}
\Pi(q^2) = i \int d^4 x e^{iqx}
\bra{0} \T\theta_\mu^\mu(x)\theta_\nu^\nu(0)\ket{0}
=
\int d\sigma^2 \frac{\rho_\theta (\sigma^2)}{\sigma^2-q^2-i\epsilon}.
\label{corr}
\end{equation}
An important theorem\cite{nsvz} for this correlator states that 
as a consequence of the broken scale invariance of QCD,
\begin{equation}
\Pi(0)=-4\bra{0} \theta_\mu^\mu(0)\ket{0}.
\label{theorem}
\end{equation}
Note that the r.h.s.\ of Eq.\ (\ref{theorem}) is divergent even in 
perturbation theory, and should therefore be regularized by
subtracting the perturbative part. The vacuum expectation value of
the $\theta_\mu^\mu$ operator then measures the energy density of 
non-perturbative
fluctuations in QCD vacuum, and the low-energy theorem (\ref{theorem})
 implies a sum rule for the spectral density:
\begin{equation}
\int \frac{d\sigma^2}{\sigma^2}
[\rho_\theta^{\rm phys}(\sigma^2)-\rho_\theta ^{\rm pt}(\sigma^2)]
=-4 \bra{0}\theta_\mu^\mu(0)\ket{0}
=-16 \epsilon_{\rm vac}
\ne 0 ,
\label{let}
\end{equation}
where the estimate for the vacuum energy density extracted from the 
sum rule analysis gives  
$\epsilon_{\rm vac}\simeq -(0.24 \mbox{ GeV})^4$ \ \cite{SVZ}.
Since the physical spectral density, $\rho_\theta^{\rm phys}$, should
approach the perturbative one,  $\rho_\theta^{\rm pt}$, at high
$\sigma^2$, the integral in Eq.\ (\ref{let}) can accumulate its value
required by the r.h.s.\ only in the region of relatively small
$\sigma^2$.

In addition,  another sum rule \cite{SVZ,Sakurai,BjK}, 
\begin{equation}
\int d\sigma^2 \rho_\theta^{\rm phys}(\sigma^2)
=
\int d\sigma^2 \rho_\theta^{\rm pt}(\sigma^2)
\label{duality}
\end{equation}
is implied by the quark--hadron duality.

\subsection{Matching onto the chiral theory}

At small invariant masses, the physical spectral density of the
correlator (\ref{corr}) should be saturated by the lightest state
allowed in the scalar channel --- two pions: 
\begin{eqnarray}
\rho_\theta^{\pi\pi} (q^2) 
& =  &
\sum (2\pi)^3 \delta^4 (p_1+p_2-q) 
| \bra{\pi(p_1) \pi(p_2)} \theta^\mu_\mu 
\ket{0}|^2 , \label{physpi}
\end{eqnarray} 
where, just as in Eq.~(\ref{pdens1}), the phase--space integral is
understood.

Since, according to Eq.~(\ref{trace}), $\theta^\mu_\mu$ is a gluonic
operator, the evaluation of Eq.~(\ref{physpi}) requires the knowledge
of the coupling of gluons to pions. This is a purely non--perturbative
problem. Nevertheless it can be rigorously solved, as it was shown
in Ref.~\cite{VZ} (see also \cite{NS}). The idea is the
following: at small pion momenta, the energy--momentum tensor can be
accurately computed using the low--energy chiral Lagrangian, 
\begin{equation}
{\mathcal{L}} = {\frac{f_{\pi}^2}{4}}\ 
\tr\ \partial_{\mu} U \partial^{\mu}U^{\dagger}
 \ +\ \frac{1}{4}\ m_\pi^2 f_\pi^2 \ \tr \left(U + U^{\dagger} \right),
\label{nlsig}
\end{equation}
where $U = \exp\left( 2i \pi / f_{\pi} \right)$,  
$\pi \equiv \pi^a T^a$ and $T^a$ are the $SU(2)$ generators
normalized by $\tr\ T^a T^b = \frac{1}{2} \delta^{ab}$.
The trace of the energy--momentum tensor for this Lagrangian is
\begin{equation}
\theta_\mu^\mu =
 - 2\ \frac{f_{\pi}^2}{4}\ \tr\ \partial_{\mu} U \partial^{\mu} 
U^{\dagger} \ - \ m_\pi^2 f_\pi^2 \ \tr \left( U + U^{\dagger} 
\right). \label{trnl}
\end{equation}
Expanding this expression (\ref{trnl}) in powers of the pion field,
one obtains, to the lowest order, 
\begin{equation}
\theta_\mu^\mu =
-\partial_\mu \pi^a \partial^\mu\pi^a +2 m_\pi^2 \pi^a \pi^a + \cdots ,
\label{trl}
\end{equation}
and this leads to an elegant result \cite{VZ} in the chiral limit of
vanishing pion mass: 
\begin{equation}
\bra{\pi^+\pi^-} \theta_\mu^\mu \ket {0} = q^2 \ .
\label{vz}
\end{equation}
This result for the
coupling of the operator $\theta_\mu^\mu$ to two pions can be
immediately generalized for any (even) number of pions using
Eq.~(\ref{trnl}).

Now that we know the coupling of gluons to the two-pion state, the
pion--pair contribution to the spectral density (\ref{physpi}) can be
easily computed by performing the simple phase space integration with
the result    
\begin{equation}
\rho^{\pi\pi}_\theta(q^2) = \frac{3}{32\pi^2}\ q^4 ; \label{ppi}
\end{equation}  
in the general case of $N_f$ light flavors, the coefficient $3$ in
Eq.~(\ref{ppi}) should be replaced by $(N_f^2-1)$. Again, the $q^4$
dependence comes only from dimensionality. Multi--pion contributions
can be evaluated using Eq.~(\ref{trnl}); we have found that at small
invariant masses their influence is small. The dominant contribution at
small invariant masses $\sigma$, in which we are primarily interested
here, therefore comes from the $\pi \pi$ state.

Recalling that to the leading order in OPE the scattering amplitude 
is dominated by the operator $\frac{1}{2} g^2 {\bf E}^{a2}$,  
we need to evaluate also the matrix element  of
the second term in Eq.~(\ref{e2}), 
$\bra{0} g^2 \theta_{00}^{(G)} \ket{\pi\pi}$  
to complete our derivation of the scattering amplitude.
As we mentioned in the previous section, this tensor operator
contributes a substantial fraction, $8/23$, to the full perturbative
result.  However, unlike
the scalar operator, the tensor term is not coupled
to the anomaly.
The contribution $\bra{0} g^2 \theta_{00}^{(G)} \ket{\pi\pi}$
therefore is of $O(g^2)$, and is sub--leading in
the heavy quark limit. In this limit, we thus come to the following
low--energy expression \cite{VZ},    
\begin{equation}
\bra{\pi \pi}\frac{1}{2}g^2 {{\bf E}^a}^2 \ket{0}  = 
\left ( \frac{4 \pi^2}{b} \right ) q^2  + O(\alpha_s, m_\pi^2). 
\label{e2pi}
\end{equation}
The matrix element in question is therefore known up to $\alpha_s$
and $m_\pi^2$ corrections.

The most important correction due to the finite pion mass is the phase
space threshold; to take it into account, we modify the spectral
density in the following way ($q^2 \geq 4m_{\pi}^2$): 
\begin{eqnarray}
\rho^{\pi\pi}_{\theta}(q^2) &=&
\frac{3}{32\pi^2} \left ( \frac{q^2-4m_\pi^2}{q^2} \right )^{1/2} 
q^4;  \label{pimass}
\end{eqnarray}
this expression should be valid at small $q^2$. Substituting this
spectral density into the general expression (\ref{yukawa}), we get
the potential due to the $\pi\pi$ exchange; at large $R$ 
\begin{equation}
V^{\pi \pi}(R) 
\rightarrow 
-\Big (\bar d_2 \frac{a_0^2}{\epsilon_0}\Big )^2
 \left ( \frac{4 \pi^2}{b} \right )^2
\frac{3}{2} (2 m_\pi)^4\  
\frac{m_\pi^{1/2}}{(4\pi R) ^ {5/2} }\  
e^{-2m_\pi R}.
\label{uncor}
\end{equation}
Note that this potential is {\it not} of Yukawa form. 
The same $R$-dependence of $\pi\pi$ exchange at large distances was
found long time ago by L{\'{e}}vy \cite{levy} and Klein \cite{klein}. 
It has been given previously also by Bhanot and Peskin~\cite{Peskin}, 
but up to an unknown constant. 
In our approach, the strength of the potential, as well as its 
dependence on the numbers of colors $N$ and flavors $N_f$ -- 
$\sim (N_f^2-1)/(11N_c - 2N_f)^2$, is fixed by 
the low-energy QCD theorems.

Note also that, unlike the perturbative result (\ref{casimir}) which is
manifestly $O(g^4)$ (besides a factor $(\bar d_2 a_0^2/\epsilon_0)^2$),
the amplitude (\ref{uncor}) is $O(g^0)$ -- 
this ``anomalously" strong interaction is the consequence of scale
anomaly\footnote{Of course, in the heavy quark limit the amplitude
(\ref{uncor}) will nevertheless vanish, since $a_0 \to 0$ and
$\epsilon_0 \to \infty$.}.

\subsection{Dynamical enhancements in the spectral density}

\begin{figure}[tb]
\centerline{%
\epsfxsize=0.45\textwidth
\epsffile{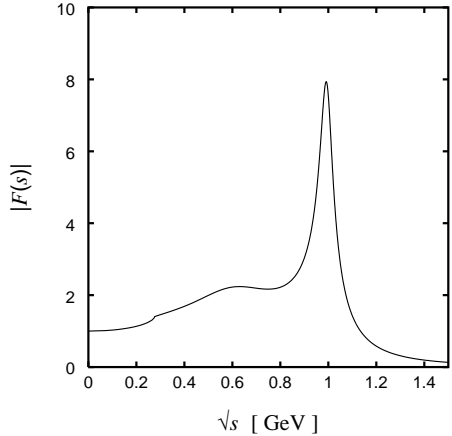}}
\caption{Scalar-isoscalar formfactor of the pion (\ref{omnes}).}
\label{fig2}
\end{figure}

The low--energy theorems\cite{nsvz,VZ} not only allow us to evaluate
explicitly the contribution of uncorrelated $\pi\pi$ exchange; they
also tell us that this contribution alone is not the complete answer 
yet. Indeed, the numerical analysis shows that the $\pi\pi$ spectral density 
(\ref{pimass}) alone cannot
saturate the sum rule (\ref{let}) -- at large $\sigma^2$, the physical
spectral density approaches the spectral density of perturbation
theory, so the integral in Eq.\ (\ref{let}) does not get any
contribution; at small $\sigma^2$, the $\pi\pi$ spectral density
(\ref{pimass}), 
according to the chiral and scale symmetries is suppressed by $\sim
\sigma^4$. The low energy theorems thus {\it require} the presence of
resonant enhancement(s) \cite{MS} in the $0^{++}$ $\pi\pi$, and
perhaps multi-pion, $\bar{K}K$ and $\eta\eta$ channels as well. Here
we will leave the complete multi-channel problem for future investigations, 
and study only the influence of these resonances in the $\pi\pi$
channel on the potential between the color dipoles.

To do this, we define the pion scalar form factor by
$\bra{\pi^+\pi^-}\theta_\mu^\mu\ket{0}=q^2 F(q^2)$ (in the chiral
limit) and write down the spectral density as  
\begin{eqnarray}
\rho_{\theta}^{\pi\pi}(s) &=&
\frac{3}{32\pi^2} \left ( \frac{s-4m_\pi^2}{s} \right )^{1/2} 
s^2 |F(s)|^2. \label{densreal}
\end{eqnarray}

It may be illustrative to consider first the idealized case of a sharp
$\sigma$ resonance. For simplicity, let us assume that the difference
between the physical and perturbative spectral densities is due to
this $\sigma$ resonance alone, and write the spectral density as
$\rho_{\theta}^{\rm phys}(s)-\rho_{\theta}^{\rm pt}(s)
=c\ \delta(s - m_\sigma^2)$.
The LET (\ref{let}) then fixes the
contribution of the narrow $\sigma$ state of mass $m_\sigma$ as 
\begin{equation}
\int \frac{ds}{s} (\rho_\theta^{\rm phys}(s)-\rho_\theta^{\rm pt}(s))
= \frac{c}{m_\sigma^2}
=-16\ \epsilon_{\rm vac} \ .
\end{equation}
The corresponding potential is of Yukawa type,
\begin{eqnarray}
V(R)&=& 
-\Big (\bar d_2 \frac{a_0^2}{\epsilon_0}\Big )^2 
\left ( \frac{4 \pi^2}{b} \right )^2
c\ \frac{1}{4\pi R}\ e^{-m_\sigma R}.
\end{eqnarray}
In this idealized situation, the strength of the potential
is directly related to the energy density of non--perturbative QCD 
vacuum.
Note, however, that this simplified model of the sharp $\sigma$
resonance is inconsistent with the asymptotics derived from the  
broken chiral symmetry ({\it Cf.}
Eq. (\ref{uncor})).

\begin{figure}[tb]
\begin{minipage}{0.45\textwidth}
\epsfxsize=\textwidth
\epsffile{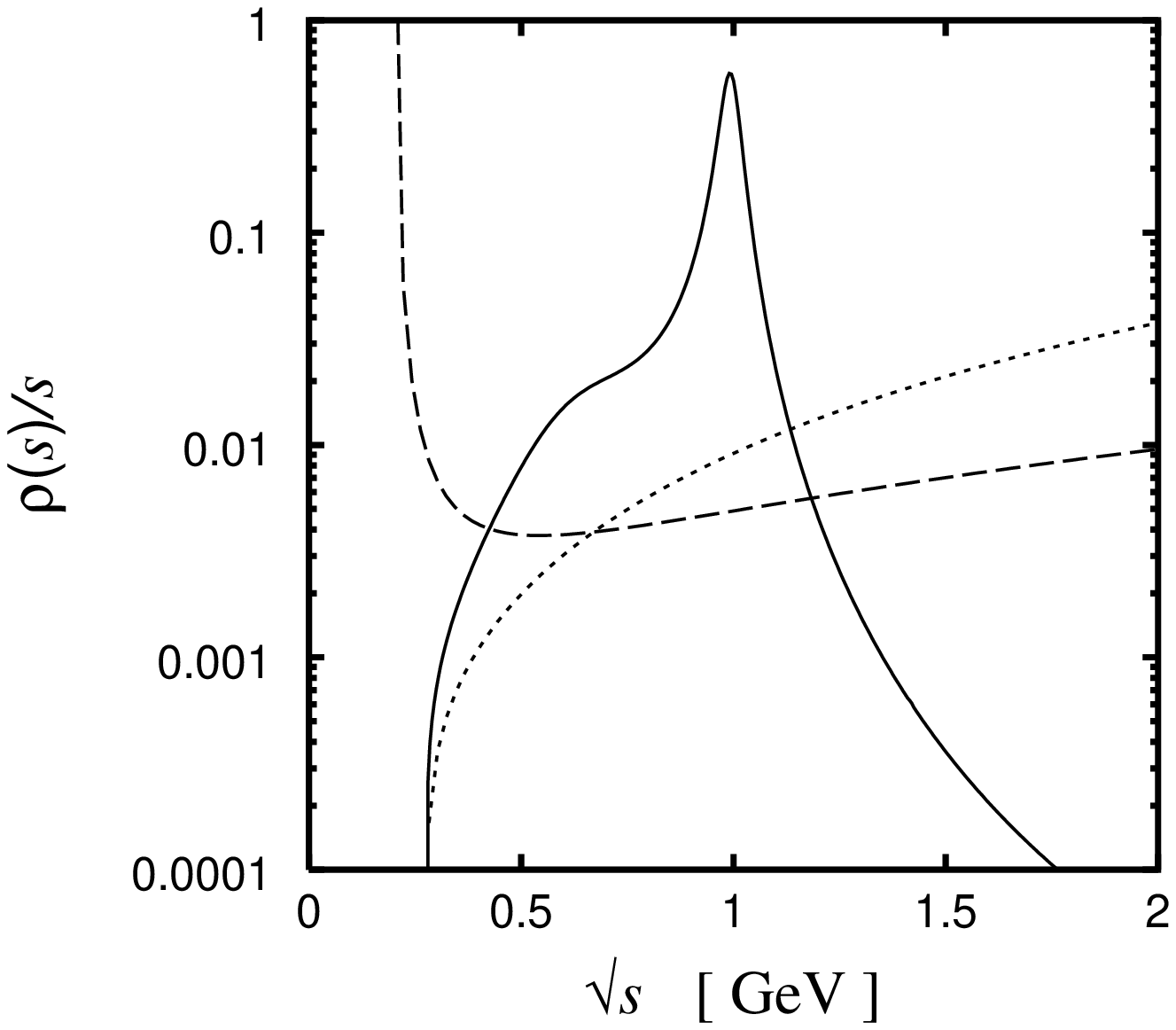}
\caption{Spectral density of the correlator 
$\bra{0}\T\theta_\mu^\mu(x) \theta_\nu^\nu(0)\ket{0}$ at 
low energy (solid line). 
The uncorrelated two--pion contribution is shown in dotted line, and
the perturbative one with one-loop running coupling constant 
($\Lambda_{\rm QCD}=200$ MeV) in dashed line.}
\label{fig3}
\end{minipage}
\hfill
\begin{minipage}{0.45\textwidth}
\epsfxsize=\textwidth
\epsffile{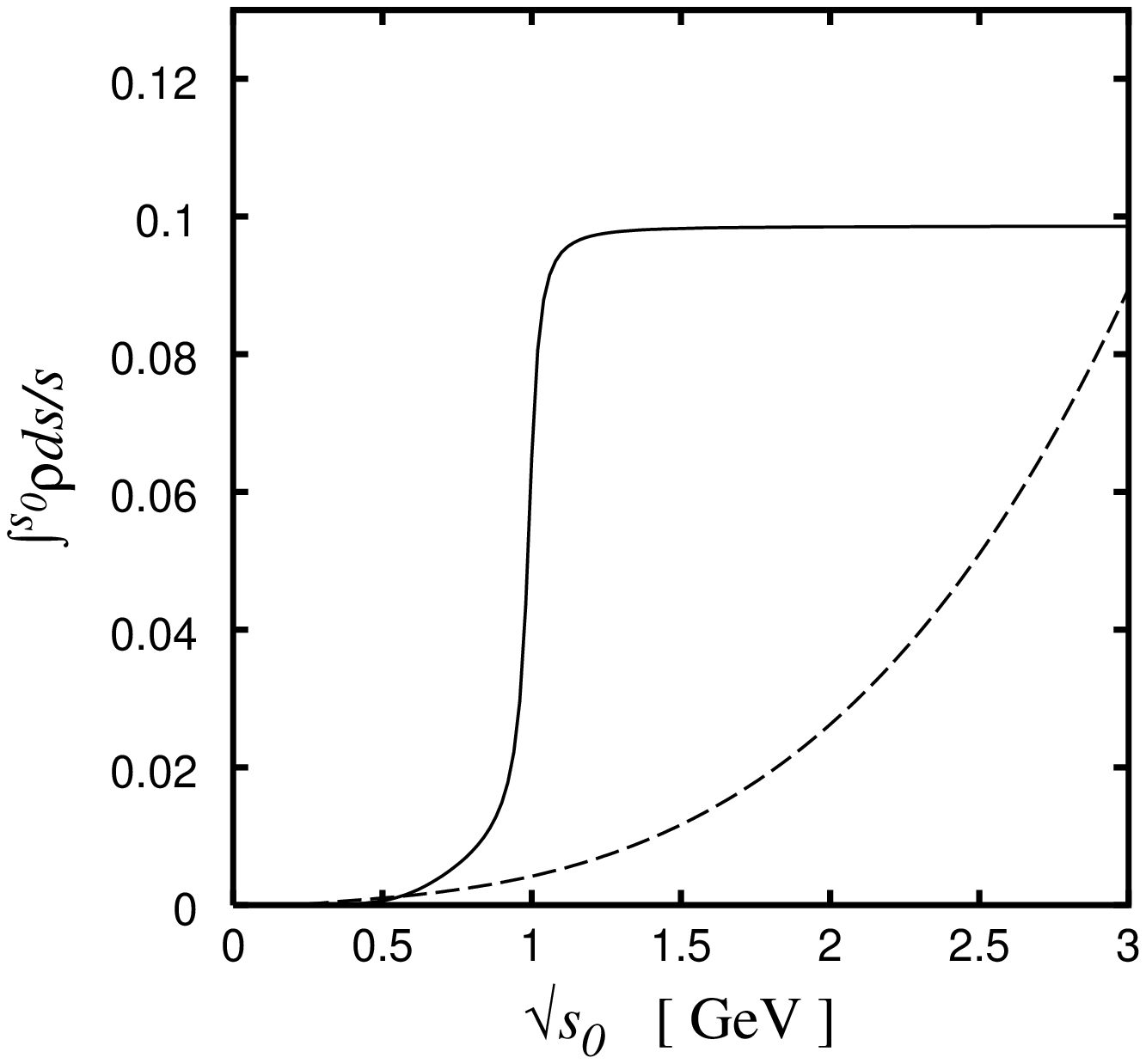}
\caption{Physical (solid) and perturbative (dashed) parts of the
integral (\ref{let})  
as a function of the upper limit, $s_0$. The LET states that the
difference of the two should be equal to the QCD vacuum energy density,
$16\ |\epsilon_{\rm vac}| \simeq 0.053$ GeV$^4$.}
\label{fig4}
\end{minipage}
\end{figure}

The formfactor $F(s)$ is directly related to the
experimental $\pi\pi$  phase shifts by the Omn\`es--Muskhelishvili
equation \cite{omnes,mus}. 
Within the single--channel treatment $F(s)$ has a solution, 
\begin{equation}
F(s) = 
\exp \left [\frac{s}{\pi}
\int_{4m_\pi^2}^{s_1} ds' 
\frac{\delta_0^0(s')}{s'(s'-s-i\epsilon)}\right ] \ ,
\label{omnes}
\end{equation}
where $\delta_0^0(s)$ is the phase shift of the $\pi\pi$ scattering in
the scalar-isoscalar channel, and formally $s_1 \to \infty$.
With this formula we can make a full use of the experimental
information on the $\pi\pi$ correlations. 

In our calculation we use a simple analytic form~\cite{Ishida} for
the phase shift $\delta_0^0(s)$ which has been shown to fit the
experimental data up to $s_{\pi\pi} \simeq 1\ {\rm GeV}^2$. Beyond this
energy, one should take into account the contributions of other
channels, such as $\bar K K$. 
We performed the integral in Eq. (\ref{omnes}) numerically up to 
$s_1$=(5 GeV)$^2$ by extrapolating the low-energy fit of the phase
shift. 
When we change $s_1$ to (20 GeV)$^2$, the
change in $F(s)$ at 1 GeV$^2$ is a few percent.
In Fig.~\ref{fig2} we show the resulting scalar formfactor of the
pion, $F(s)$. The structure of $F(s)$ may be interpreted as due to  
a broad $\sigma$ and narrow $f_0$ resonances.
For a more realistic evaluation of the formfactor, the multi--channel
calculation has to be done; the results will be reported elsewhere.

In this paper, as a simple model for the 
$\rho_\theta^{\rm phys}$, we will take the form   
\begin{eqnarray}
\rho^{\rm phys}_\theta(s)
=\left \{ 
\begin{array}{cc}
\rho^{\pi\pi}_\theta(s) & \qquad (4m_\pi^2 < s < s_0) ,\\
\rho^{\rm pt}_\theta(s) & \qquad (s_0 < s) ,
\end{array} 
\right.    
\label{specmodel}
\end{eqnarray}
where $s_0$ is a matching scale.

\subsection{The analysis of the sum rule}

Let us consider the sum rule (\ref{let})
within our simple model for the spectral density.
When the model (\ref{specmodel}) for $\rho^{\rm phys}_\theta(s)$ is
used, the upper limit of the integral in Eq. (\ref{let}) can be
replaced by $s_0$.
In Fig.~\ref{fig3} we show the physical and perturbative parts of the
integrand in the sum rule (\ref{let}) with solid and dashed lines,
respectively. 
Since for the spectral density in the perturbation theory there is
no scale other than $s$, the coupling constant should be taken running
with this scale:
\begin{eqnarray}
\rho_\theta^{\rm pt}(s) 
&=&
\left ( \frac{9\alpha_s (s)}{8\pi}\right )^2 \frac{2}{\pi^2}\ s^2 ,
\label{running}
\end{eqnarray}
where $\alpha_s(s)= 4\pi/(b\ln(s/\Lambda_{\rm QCD}^2))$ with 
$\Lambda_{\rm QCD}$=200 MeV.

We note that the spectral density for uncorrelated pions 
$\rho^{\pi\pi}_\theta$
(\ref{pimass}), which is shown in dotted line in Fig.~\ref{fig3},
 has the same functional
form as the $\rho^{\rm pt}_\theta$, up to the logarithm and the
threshold factor. As a consequence of this, we find that the
uncorrelated $\pi\pi$ spectral density (\ref{pimass}) 
cannot obey both the sum rule (\ref{let}) and the duality constraint 
(\ref{duality}).

\begin{figure}[t]
\centerline{\epsfxsize=0.45\textwidth
\epsffile{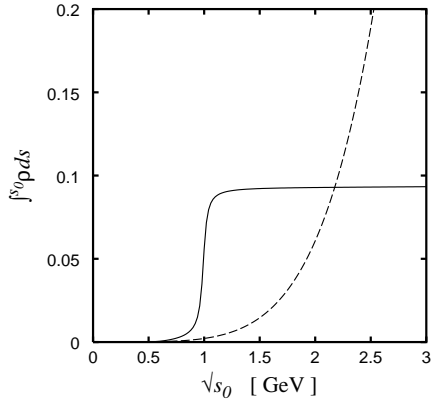}}
\caption{Integral of the duality relation. The notations are the same
as in Fig.\ \ref{fig4}.} 
\label{fig5}
\end{figure}

On the other hand, the spectral density
obtained with the Omn\`es--Muskhelishvili solution has a non-trivial
structure (see Fig.~\ref{fig3});
one can clearly see a narrow peak at the $f_0$ resonance region with
a shoulder coming from the broad ``$\sigma$'' around 0.6 GeV. 

We plot the integral in the LET for the physical (solid) and
perturbative (dashed) parts separately as a function of the upper
limit, $s_0$. 
One can see that the value of the integral for the physical spectral
density is mainly accumulated in 1 GeV region, and the ``$\sigma$''
contributes to it about 20 \% (see Fig.~\ref{fig4}). The perturbative
part behaves as $s^2$ up to the logarithm, weighting the higher energy
region. The LET tells us that the difference of these two
contributions should be equal to the energy density of the QCD vacuum.
In our model for the spectral density, the LET (\ref{let}) is 
satisfied when we choose  $s_0$=(2$\sim$2.5 GeV)$^2$. 
As for the duality relation (\ref{duality}), the equality of the
integrals of the physical and perturbative spectral densities is
achieved when we choose $s_0 \sim (2\ \rm{GeV})^2$ (Fig.~{fig5})
 -- this value of 
the matching scale therefore provides a consistent solution to both 
the LET and the duality relation.  

Even though our spectral density cannot be taken seriously
in the high mass region beyond $\sim$1 GeV, our calculation nevertheless 
shows the following:
Non-perturbative dynamics of QCD generates 
enhancements in the intermediate mass region in the form of 
hadronic resonances,
which make the physical spectral density consistent with the LET (\ref{let}). 
The narrow $f_0(980)$ is more important for the LET than the low mass,
broad $\sigma$ resonance. Therefore, to discuss the influence of heavier 
resonances (like $f_0(1500)$) we need to perform a 
coupled--channel analysis including the $\bar KK$ and other states.
In the rest of this paper we put the matching scale $\sqrt{s_0}$ = 2 GeV.

\section{The potential between color dipoles}
\label{jpsipot}

As a concrete example, let us consider the potential
between two $J/\psi$'s at rest.
Although the charm quark is perhaps not heavy enough to justify 
the heavy quark limit, we
try to extrapolate our result to $J/\psi$ and discuss its
implications.

For the pure Coulombic bound state, $\bar d_2 = 7/36$, and
the Bohr radius and Rydberg energy are given by 
$a_0=4/(3\alpha_s m)$ and $\epsilon_0 = (3 \alpha_s/4)^2 m=1/(a_0^2 m)$,
respectively ($\alpha_s=g^2/4\pi$). 
We have $\alpha_s(J/\psi)$=0.87 and $a_0$=0.20 fm for the $J/\psi$ 
with the phenomenological inputs,
$\epsilon_0 = 2 M_D-M(J/\psi)$=642 MeV and $m$=1.5 GeV.
These values show the application to $J/\psi$ will be qualitative at
most, because of the large $\alpha_s$ value and because  
$a_0 > s_0^{-1/2}$; the latter means that nonperturbative effects
penetrate inside the radius of $J/\psi$.

\begin{figure}[t]
\centerline{\epsfxsize=0.45\textwidth
\epsffile{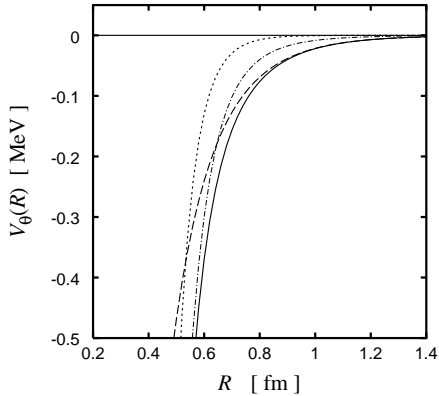}} 
\caption{Potential (\ref{yukawa}) between two $J/\psi$'s (solid
line). Contributions of the spectral densities of $s>s_0$ and $s<s_0$
with $\sqrt{s_0}=2$ GeV,
respectively, are plotted in dotted and dashed lines. The perturbative
result (\ref{casimir}) is shown in dashed-dotted line, for reference.}
\label{fig-pot}
\end{figure}

In Fig.~\ref{fig-pot} the resulting potential (\ref{yukawa})
between two $J/\psi$'s is shown as a solid line. 
In our model for the spectral density, the potential consists of two
components, high $q^2$ (dotted) and low $q^2$ (dashed), separated by
$s_0$, which we set (2 GeV)$^2$.  (As in the heavy quark 
limit, we omit the
contribution from the tensor exchange, $\theta_{00}^{(G)}$).

First, we see that the potential at large distances is naturally  
determined by the spectral density of the low $q^2$ region.  Moreover
the total strength of the potential at large distances is enhanced by the
non-perturbative spectrum of QCD, compared to the formal
perturbative result (\ref{casimir}) denoted by the dashed--dotted line. The
region where the two components compete is $R \simeq 0.5\sim0.6$ fm, 
which is 
much larger than the scale determined by $s_0^{-1/2} \sim 0.1$ fm.  
This is in contrast with naive expectation that beyond the scale 
$s_0^{-1/2}$, the potential should be dominated by the non--perturbative 
spectral density.
The reason for this lies in the
large value of $\alpha_s(J/\psi)$, reflecting the fact that charmonium is 
still far from the heavy quark limit.

In the discussion of the LET (\ref{let}), we used the running coupling
constant, while the coupling constant used
here is frozen at the $J/\psi$ scale. This is because in the 
heavy quark limit,
it is natural to renormalize the coupling constant at the scale of 
quarkonium, $g(\epsilon_0)$, with $\epsilon_0 \gg s_0^{1/2},
\Lambda_{\rm QCD}$. The matrix element of $G^2$ should then contain
the effects of quantum fluctuations below this energy scale. Within
the perturbative approach the renormalization group 
ensures independence of the final result on the choice of  
renormalization point, at least in the leading-log approximation, 
which we used in Eq.~(\ref{running}). 
In the case of $J/\psi$, the renormalization scale (chosen at the
binding  energy, $\epsilon_0$) is  
still {\it lower} than $s_0$, and the spectral density (\ref{pdens2})
with fixed $\alpha_s(J/\psi)$ is significantly larger than the one
with the running coupling constant (\ref{running}). Again, this
reflects the fact that non-perturbative effects penetrate ``inside"
the $J/\psi$. The most important feature seen in Fig.~\ref{fig-pot} is
the dominance of low-$q^2$ enhancements in the spectral density in the
behavior of potential at large distances.

\section{The sum rule for the potential}
\label{sec:sumrule}

We can derive
an interesting sum rule for the strength of the potential; according
to Eqs.~(\ref{yukawa}) and (\ref{casimir}), we have 
\begin{eqnarray}
\int_a^\infty d^3{\bf R} 
  \big ( V_\theta(R)-V_\theta^{\rm pt}(R)\big ) =
- \Big ( \bar d_2 \frac{a_0^2}{\epsilon_0} \Big )^2
  \Big ( \frac{4\pi^2}{b} \Big )^2
\int \frac{d\sigma^2 }{\sigma^2} 
\big (
 \rho_\theta^{\rm phys}(\sigma^2)-\rho_\theta^{\rm pt}(\sigma^2)
\big )
\Gamma(2,\sigma a) \ ,
\label{potsr1}
\end{eqnarray}
where $a$ should be chosen to be of the order of the onium radius, and
$\Gamma(z,p)=\int_p^\infty dt t^{z-1}e^{-t}$.  
As we discussed previously, the physical spectral density 
$\rho^{\rm phys}_\theta(\sigma^2)$ differs from the perturbative one,
$\rho^{\rm pt}_\theta(\sigma^2)$, in the region $\sigma^2 \lesssim s_0$.
In the heavy quark limit, $a \propto 1/(\alpha_s m)$ and 
$a \sqrt{s_0} \ll 1$.
Therefore we can re--write the sum rule (\ref{potsr1}) in a more 
suggestive form: 
\begin{eqnarray}
\int_a^\infty d^3{\bf R} \big ( V_\theta(R)-V_\theta^{\rm pt}(R)\big )
& =&
- \Big ( \bar d_2 \frac{a_0^2}{\epsilon_0} \Big )^2
  \Big ( \frac{4\pi^2}{b} \Big )^2
\int \frac{d\sigma^2 }{\sigma^2} 
\big (
 \rho_\theta^{\rm phys}(\sigma^2)-\rho_\theta^{\rm pt}(\sigma^2)
\big ) 
\nonumber \\
&=&
- \Big ( \bar d_2 \frac{a_0^2}{\epsilon_0} \Big )^2
  \Big ( \frac{4\pi^2}{b} \Big )^2
  16|\epsilon_{\rm vac}| \ ,
\label{potsr2}
\end{eqnarray}
which relates the overall strength of the interaction between small 
color dipoles to the energy density of the non-perturbative 
QCD vacuum.

\section{Quarkonium interactions with pions}
\label{sec:pion}

As another application of our formalism, we evaluate the 
cross sections of elastic scattering $\pi \Phi \to \pi
\Phi$ and of excitation process $\pi \Phi \to \pi \Phi'$; the latter
cross section was previously computed in Refs.~\cite{SSZ,CS}. 
These cross sections are important for the analyses of quarkonium 
production in heavy ion collisions \cite{MS1,QM}. 
The fact that soft pions effectively 
decouple from heavy quarkonia was previously noted in Ref \cite{DK}.

\subsection{Elastic $\pi \Phi$  scattering}

Within the OPE formalism (\ref{ope0}), it is straightforward to write
down the amplitude of pion--quarkonium elastic scattering at
small energies: to the leading order in OPE,
\begin{eqnarray}
{\cal M}^{kl}(P',p';P,p) &=& - \bar d_2 \frac{a_0^2}{\epsilon_0}
\bra{\pi^k(p')}\frac{1}{2}g^2{\bf E}^{a2}(0) \ket{\pi^l(p)}.
\end{eqnarray}
The matrix element  $\bra{\pi^k}g^2{\bf E}^{a2}(0)\ket{\pi^l}$ can be 
found from $\bra{\pi^k\pi^l}g^2{\bf E}^{a2}(0)\ket{0}$,
(\ref{e2pi}), by crossing;
the LET (\ref{e2pi}) tells us that 
up to $\alpha_s$ and $m_\pi^2$ corrections
\begin{equation}
\bra{\pi^k(p')}\frac{1}{2}g^2{\bf E}^{a2}(0)\ket{\pi^l(p)}
=\frac{4\pi^2}{b}
\bra{\pi^k(p')}\theta_\mu^\mu(0)\ket{\pi^l(p)}
=\delta^{kl} \frac{4\pi^2}{b}\ t\ F(t) ,
\end{equation}
where $t=(p-p')^2$,
and we have introduced, as before, the pion scalar formfactor, $F(t)$. 
Taking into account the non-relativistic normalization of the $\Phi$
state, we have the expression for the total elastic cross section in
the CM frame,
\begin{eqnarray}
\sigma(s)
&=&
\frac{1}{2p_0 \bar v_{rel}}
\int \frac{d^3P'}{(2\pi)^3} \int \frac{d^3p'}{(2\pi)^3 2p'_0}
|{\cal M}|^2\ (2\pi)^4 \delta^4(P'+p'-P-p)
\nonumber \\
&=&
\frac{1}{16\pi s}\frac{M^2}{{\bf p}^2}
\left ( \bar d_2 \frac{a_0^2}{\epsilon_0} \right ) ^2
\left ( \frac{4 \pi^2}{b} \right ) ^2 
\int_0^{4{\bf p}^2} d(-t)\ t^2\ |F(t)|^2 , \label{picross}
\end{eqnarray}
where $\bar v_{rel}=\sqrt{(P\cdot p)^2-M^2 m^2}/P^0 p^0$ is the
relative velocity of the incoming $J/\psi$ and pion, and
${\bf p}^2 = (s-(M-m_\pi)^2)(s-(M+m_\pi)^2)/4s$ is the CM
momentum.

\begin{figure}[tb]
\begin{minipage}{0.48\textwidth}
\epsfxsize=\textwidth
\epsffile{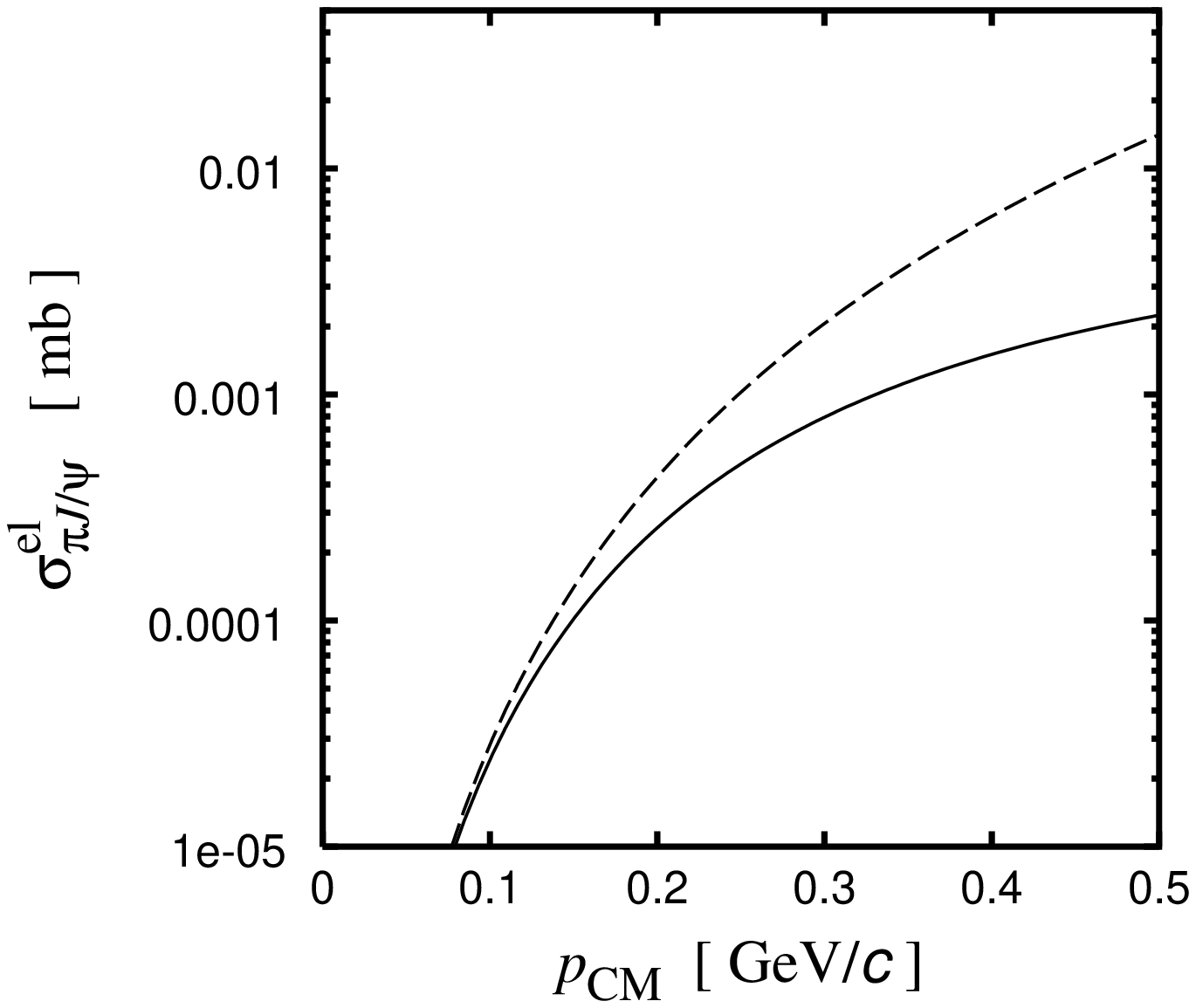}
\caption{$\pi J/\psi$ elastic cross section (solid) as a function of the
CM momentum. Dashed line is the case of $F(t)$=1.}
\label{elastic}
\end{minipage}
\hfill
\begin{minipage}{0.48\textwidth}
\epsfxsize=\textwidth
\epsffile{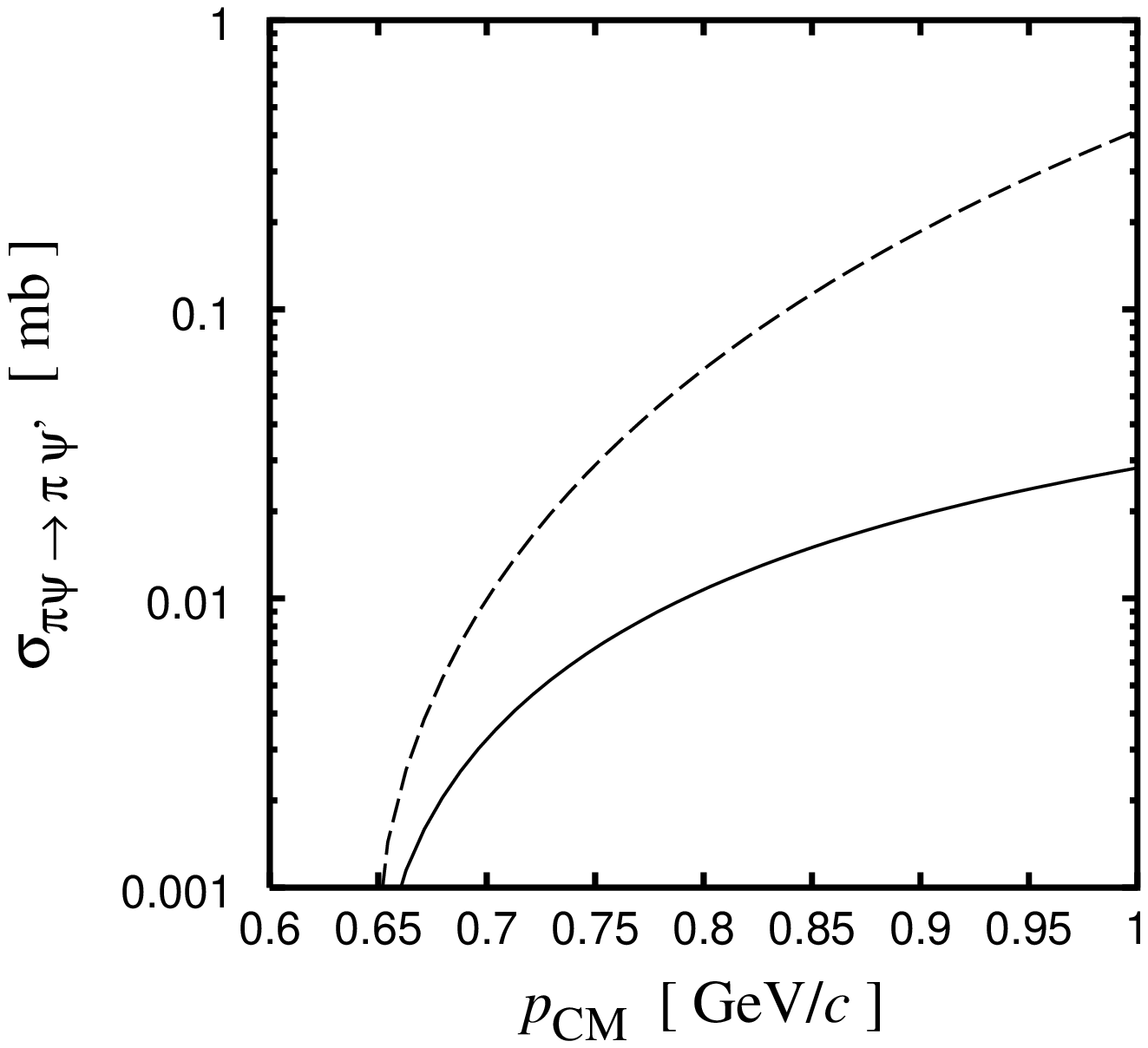}
\caption{$\pi J/\psi \to \pi \psi'$ cross section. The notations are
the same as in Fig.~\ref{elastic}.} 
\label{inelastic}
\end{minipage}
\end{figure}

The result for the elastic $\pi J/\psi$ cross section is shown in
 Fig.~\ref{elastic}. The pion scalar formfactor $F(q^2)$ and other
parameters are the same as in the previous section.  
Note that the chiral symmetry requires a 
strong momentum dependence, $t^2$ in Eq.~(\ref{picross}); therefore 
at low energies the $J/\psi$ interaction with pions is very weak. 
Extrapolation of the scalar formfactor
$F(t)$ to the scattering region, $t<0$, induces additional
suppression of the $\pi\ J/\psi$ interaction.
At small energies (see Fig.~\ref{elastic}) the cross section is on the 
order of 0.01 mb, which is much smaller than the geometrical cross
 section  of the $J/\psi$. For quarkonium production in heavy ion
 collisions, this implies that the interactions with secondary pions
 do not contribute to the broadening of the quarkonium transverse
 momentum spectra.

\subsection{$\pi \Phi \to \pi \Phi'$ transition amplitude}

Our next example is the transition process, $\pi \Phi \to \pi \Phi'$.
In this case, however, the transferred momentum is on the order of the
binding energy,
$\Delta=M'-M=O(\epsilon_0) $,
which may invalidate our assumption on the factorization between
the short and long distances. 
Fortunately, since the size of quarkonium $a_0 \sim 1/(g^2 m) 
\ll 1/\epsilon_0 \sim 1/(g^4 m)$ in the heavy quark limit, 
 the double-dipole form 
\begin{equation}
{\cal M} = \bra{\phi' \pi} \tr \Big [
{\bf r\cdot E}\frac{1}{H_a+\epsilon+iD^0}{\bf r\cdot E}
\Big ] \ket{\phi \pi}
\label{inel}
\label{ddipole'}
\end{equation}
is still valid \cite{Peskin}. 
The structure of this formula is transparent: the initial quarkonium 
$\Phi$ absorbs/emits a
gluon, then propagates with the internal energy, $-\epsilon+Q$, and
emits/absorbs another gluon to form a color-singlet, excited
quarkonium state $\Phi'$; 
these gluons originate from pions.

To apply our formalism, let us approximate $-iD^0$ in Eq.~(\ref{ddipole'})
by the typical value of the gluon momentum, $\Delta$.
Within this (rather crude) approximation, the quarkonium part and 
the pion part can be
factorized in the matrix element, (\ref{inel}), and
the relevant Wilson coefficient, which for this process reads
\begin{equation}
\bar d'_2\frac{a_0^2}{\epsilon_0}
= \frac{1}{3N} \bra{\phi'}r^i\frac{1}{H_a+\epsilon-\Delta}r^i \ket{\phi}.
\end{equation}
The transition matrix element then reduces to the same form as in the
elastic case:
\begin{eqnarray}
{\cal M}= \left ( \bar d'_2 \frac{a_0^2}{\epsilon_0}\right )
\left ( \frac{4\pi^2}{b}\right )\ t \  F(t),
\end{eqnarray}
and the total transition cross section can be written as
\begin{equation}
\sigma(s)=
\frac{1}{16\pi s}\frac{M M'}{ {\bf p}^2}
\left ( \bar d'_2 \frac{a_0^2}{\epsilon_0} \right ) ^2
\left ( \frac{4 \pi^2}{b} \right ) ^2 
\int_{t_{min}}^{t_{max}} d(-t) \ t^2 \ |F(t)|^2.
\end{equation}

In Fig.~\ref{inelastic} we show the result for the $\pi J/\psi \to
 \pi \psi'$ cross section, evaluated assuming the 1s and 2s Coulomb
 wave functions, and  $\Delta= (3/4) \epsilon_0$. It shows that the
 cross section is on the  order of 0.01-0.1 mb.   

We also evaluated the partial width of the $\psi'$ due to
$\psi' \to \psi \pi \pi$ decay within the same formalism, and 
obtained $\Gamma$=260 (70) keV with
(without) using the formfactor $F(s)$. This should be compared with
the experimental value, $135\pm 20$ keV \cite{PDG}. We conclude that
our calculations, due to the assumption of the heavy quark limit, hold
to within a factor of 2 only. Additional confirmation of the $s$
dependence of the matrix element comes from the dipion invariant mass
distribution in the $\psi' \to \psi \pi \pi$ decay \cite{spectrum}.

\section{Summary and discussion}
\label{sec:sum}

We have shown that at large distances the interaction of small 
color dipoles becomes totally non-perturbative. 
This result has a deep physical origin: indeed,   
one can trace it back to the sum rule (\ref{let})
for the correlator of the energy--momentum tensor, 
which reflects the fact that the non--perturbative 
vacuum of QCD is characterized by non--zero energy density. 

For QCD practitioners, ``non--perturbative" is often a substitute 
for ``incalculable". Nonetheless, in our case, we were able to 
evaluate explicitly this, non--perturbative, scattering amplitude 
in a model--independent way. The key ingredients in our approach 
were {\mbox {\it{ i)}} the use} of spectral representation in 
the $t-$channel and 
$ii)$ the low--energy theorem arising from 
the (broken) scale and chiral invariances of QCD. 

\begin{figure}[tb]
\centerline{\epsfxsize=0.45\textwidth
\epsffile{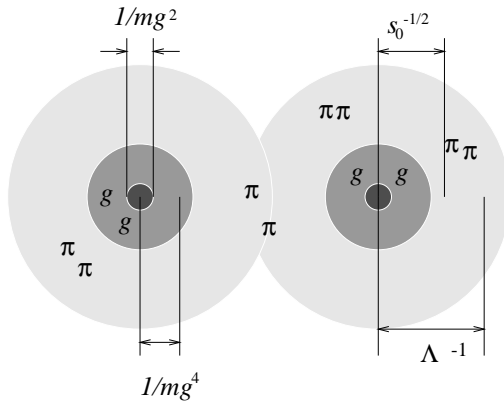}} 
\caption{Schematic picture of the potential between heavy onia.}
\label{fig-idea}
\end{figure}

What are the implications of our results? First, we find that   
 the long--distance interactions of small color 
dipoles are dominated by pion clouds; 
the qualitative picture of this interaction is illustrated in 
Fig.~\ref{fig-idea}. The size occupied by the heavy quark--antiquark 
pair in the quarkonium (see Fig.~\ref{fig-idea}) is 
$a_0 \sim 1/(g^2 m)$; the gluon cloud spreads 
up to the distances on the order of the inverse binding energy
 $1/\epsilon_0 \sim 1/(g^4 m)$, since the typical momentum $K$ of
 gluons is $K \sim \epsilon_0$. (This picture of quarkonium structure
 emerges also from the NRQCD approach of Bodwin, Braaten and Lepage
 \cite{BBL}.) The pion cloud begins to dominate at the distances 
$\sim s_0^{-1/2}$, and spreads up to the distances $(2\mu)^{-1}$, 
where $\mu$ is the pion mass ($s_0$ is the mass scale at which the 
non-perturbative effects begin to dominate in the spectral density,
 see Section III). This pion cloud may as well be 
important in high--energy scattering. One may even speculate
(see Bjorken \cite{Bjorken}) 
that pions are responsible for the so-called ``soft pomeron", 
even though it is not yet clear how to extend our calculations 
to high--energy scattering -- this would require evaluation 
of higher orders in the multipole expansion. Nevertheless the 
possibility that the diffusion of partons toward small $k_t$ 
makes pionic degrees of freedom important looks very plausible to us.

Second, the fact that pions (and therefore light quarks) dominate 
the long--distance interactions of heavy quark systems is important
for lattice QCD simulations. Even though naively one may think that 
light quarks are relatively unimportant for the studies of heavy 
quarkonia on the lattice, our findings show that the 
opposite is true. This suggests that to extract the properties 
of heavy quarkonia from the lattice QCD one has to use 
``unquenched" theory with light quarks. The importance of pionic 
degrees of freedom in determination of the mass splittings of heavy 
quarkonia was investigated in Ref \cite{GR}.

Third, we find  
that both inelastic and elastic $\pi J/\psi$ scattering cross 
sections are very small, less than $0.1$ mb. The smallness of 
inelastic cross section suggests that pions are very ineffective 
in dissociating $J/\psi$'s, lending support to the idea to use 
quarkonia as a signal of deconfinement \cite{MS1}. 
The smallness of the elastic cross 
section explains why the transverse momentum distributions of 
$J/\psi$'s seem to be unaffected by the final state interactions 
with secondary pions, whereas much bigger $\psi'$'s, to which 
our multipole expansion analysis does not apply, can show significantly 
larger mean transverse momenta.

\acknowledgements

It is a pleasure to thank J. Ellis, M. Gyulassy, T.D. Lee,
L.D. McLerran, H. Mishra, 
A.H. Mueller, R.D. Pisarski, H. Satz, E.V. Shuryak and
M.B. Voloshin for useful discussions, and M. Savage for correspondence
and for bringing Ref \cite{GR} to our attention. 


\appendix
\section{Derivation of Eqs.~(14) and (16)} 
The Feynman propagator of a scalar field $\varphi$ of mass $\sigma$ 
is defined by
\begin{eqnarray}
i \Delta_F(x;\sigma^2) &=&
i\langle \T \varphi(x) \varphi(0) \rangle
\nonumber \\
&=& i \int \frac{d^4k}{(2\pi)^3}
\delta(k^2-\sigma^2)\theta(k_0)
(e^{-ikx}\theta(x^0)+e^{ikx}\theta(-x^0))
\nonumber \\
&=& \int \frac{d^4k}{(2\pi)^4}\frac{e^{-ikx}}{\sigma^2-k^2-i\epsilon}
\nonumber \\
&=&
\frac{i}{4\pi^2}\frac{\sigma^2}{\sqrt{(-x^2+i\epsilon)\sigma^2}}
K_1\big (\sqrt{(-x^2+i\epsilon)\sigma^2}\big ) \ ,
\end{eqnarray}
where $K_1$ is the modified Bessel function.
The Born amplitude of one-boson exchange with coupling $g$ 
is
\begin{equation}
i{\cal M}_B({\bf q}) = \int d^4 x e^{i{\bf q\cdot x}}
\langle \T ig \varphi(x) ig \varphi(0) \rangle
=ig^2 \int d^4 x e^{i{\bf q\cdot x}} i\Delta_F(x;\sigma^2),
\end{equation}
which may be related to an interaction potential by 
$i{\cal M}({\bf q})=-iV({\bf q})$.
Going to Euclidean space, 
where $x_E^2= \tau^2+{\bf x}^2 = \tau^2+R^2$, we can show that
\begin{eqnarray}
V(R) =- g^2\int_{-\infty}^\infty dt i\Delta_F(x;\sigma^2)
 &=&
-g^2\int_{-\infty}^\infty d\tau \frac{1}{4\pi^2}
\frac{\sigma^2}{\sqrt{x_E^2\sigma^2}}
K_1\big (\sqrt{x_E^2\sigma^2}\big )
=-\frac{g^2}{4\pi R}e^{-\sigma R} .
\end{eqnarray}

To the leading in the OPE, the potential between color dipoles
(\ref{yukawa}) is 
a superposition of this Yukawa potential with the spectral function,
$\rho_\theta(\sigma^2)$. 
In the perturbative calculation, the matrix element of $G^2$ between
the vacuum and the two-gluon state is
\begin{eqnarray}
\bra{p_1 \varepsilon_1 a , p_2 \varepsilon_2 b}G^{\ab c}G_{\ab}^c \ket{0} 
&=& 4 (-p_1 \cdot p_2 \ \varepsilon^*_1\cdot \varepsilon^*_2
       +p_1 \cdot \varepsilon^*_2 \ p_2\cdot \varepsilon^*_1 )
\delta^{ab} +O(g^2),
\end{eqnarray}
where $p_i$, $\varepsilon_i$ and $a$ are the momentum, polarization
and $SU(N)$ color index of the gluon, respectively.
Noting that the sum over physical polarizations yields the projection,
$
\sum_{\rm pol}\varepsilon_{\mu}\varepsilon^*_{\nu}
=\delta_{ij}-p_ip_j/{\bf p}^2,
$
we have a compact expression,
\begin{eqnarray}
\sum_{\rm pol,col} |\bra{p_1 \varepsilon_1 a , p_2 \varepsilon_2 b}
 G^2 \ket{0} |^2 
= 
4^2 (N^2-1) \ 2 (p_1 \cdot p_2)^2
=   8(N^2-1) q^4 ,
\end{eqnarray}
which depends only on $q^2=(p_1+p_2)^2$.
With the phase space factor of two identical particles,
\begin{eqnarray}
\frac{1}{2} \int \frac{d^3p_1}{(2\pi)^3 2\omega_1}
\int \frac{d^3p_2}{(2\pi)^3 2\omega_2}
(2\pi)^3 \delta^4(p_1+p_2-q) 
= \frac{1}{32\pi^2},
\end{eqnarray}
we find Eq.~(\ref{pdens2}) as the spectral density of the correlator of
$\theta_\mu^\mu = -(b g^2/32 \pi^2) G^2$, to the leading in $g$.

Similarly the spectral density of the two-pion states (\ref{ppi}) can
be calculated, but with 
$\bra{\pi^k\pi^l} \theta_\mu^\mu \ket{0}=q^2\delta^{kl}$
in the chiral limit ($k,l=1,2,3$).

\section{Alternative derivation of Eq.~(17)}

To confirm our result, we derive here Eq. (\ref{casimir}) applying  
the functional method of Ref.~\cite{Peskin} to the
scalar part of the interaction (\ref{pot1}) (in
Euclidean space);
\begin{eqnarray}
V_\theta(R) &=&
- \Big ( \bar d_2 \frac{a_0^2}{\epsilon_0}\Big )^2 
\int_{-\infty}^\infty d\tau  \bra{0} 
\frac{g^2}{8} G^2(x) \frac{g^2}{8} G^2(0)\ket{0}
\nonumber \\
&=& 
- \Big ( \bar d_2 \frac{a_0^2}{\epsilon_0}\Big )^2 
 \frac{g^4}{32} \int_{-\infty}^\infty  d\tau
\bra{0}  G^a_{\alpha\beta}(x) G^b_{\alpha'\beta'}(0)\ket{0}^2,
\label{pesk}
\end{eqnarray}
where all indices are summed over.
Using the expression for the two--point function of the gluon in 
Feynman gauge, 
\begin{eqnarray}
\langle A_\mu^a(x) A_\nu^b(y)\rangle &=&
\frac{1}{4\pi^2}\frac{1}{(x-y)^2}\delta_{\mn}\delta^{ab} ,
\end{eqnarray}
we have
\begin{eqnarray}
\langle G^a_{\mu\nu}(x)G^b_{\mu'\nu'}(0)\rangle &=&
\frac{2}{4\pi^2} \frac{\delta^{ab}}{x^6}
\left \{
  \delta_{\nu\nu '}  (\delta_{\mu\mu '}x^2 - 4x_\mu x_{\mu '})
 + \delta_{\mu\mu'} (\delta_{\nu\nu'}x^2 - 4x_\nu x_{\nu'}) \right .
\nonumber \\ &&
\left .
- \delta_{\nu\mu'} (\delta_{\mu\nu'}x^2 - 4x_\mu x_{\nu'})
- \delta_{\mu\nu'} (\delta_{\nu\mu'}x^2 - 4x_\nu x_{\mu'})
\right \} +O(g^2),
\end{eqnarray}
and
\begin{eqnarray}
\langle G^a_{\mu\nu}(x)G^b_{\mu'\nu'}(0)\rangle ^2
&=&
(N^2-1)\frac{24}{\pi^4}\frac{1}{x^8}+O(g^2).
\label{2point2}
\end{eqnarray}
Substituting Eq.~(\ref{2point2}) into Eq.~(\ref{pesk}), 
we obtain the leading expression (\ref{casimir}) for the potential of the scalar part
($x^2=\tau^2+R^2$),
\begin{eqnarray}
V_\theta(R)
&=& 
- \Big ( \bar d_2 \frac{a_0^2}{\epsilon_0}\Big )^2 
\frac{g^4}{32}
\int_{-\infty}^\infty d\tau  
(N^2-1)\frac{24}{\pi^4}\frac{1}{x^8},
\nonumber \\
&=& 
- g^4 \Big (\bar d_2 \frac{a_0^2}{\epsilon_0}\Big )^2 
\frac{15}{8\pi^3}\frac{1}{R^7}.
\end{eqnarray}

\end{document}